\documentclass[10pt,a4paper,english]{article}
        \usepackage[T1]{fontenc}
        \usepackage[utf8]{inputenc}
        \usepackage{babel}
        \usepackage{array} 
        \usepackage{slashbox}
        \usepackage{booktabs}
        \usepackage{etextools}
        \usepackage{subfig}
        \usepackage{framed}
        \usepackage{listings}
        \usepackage[bottom]{footmisc}
        \usepackage{scalerel,stackengine}
        \usepackage{multirow}

\stackMath
\newcommand\reallywidehat[1]{%
\savestack{\tmpbox}{\stretchto{%
  \scaleto{%
    \scalerel*[\widthof{\ensuremath{#1}}]{\kern-.6pt\bigwedge\kern-.6pt}%
    {\rule[-\textheight/2]{1ex}{\textheight}}
  }{\textheight}%
}{0.5ex}}%
\stackon[1pt]{#1}{\tmpbox}%
}
        \usepackage{setspace,graphicx,epstopdf,amsmath,amsfonts,amssymb,amsthm}
\usepackage{float}
\usepackage{verbatim,amscd, mathabx}
\usepackage[titletoc]{appendix}

\newtheorem{theorem}{Theorem}[section]

\newtheorem{proposition}[theorem]{Proposition}

\numberwithin{equation}{section}

\newcommand{\K}{\mathbb{K}}

\renewcommand{\K}{\mathcal{K}}

\renewcommand{\d}{\ensuremath{\operatorname{d}\!}}

\makeatletter
\def\blfootnote{\xdef\@thefnmark{}\@footnotetext}
\makeatother

\begin{document}

\title{Matching distributions:\\
Recovery of implied physical densities\\ from option prices}


\author{Jarno Talponen}

\date{\today}              


\renewcommand{\thefootnote}{\fnsymbol{footnote}}

\singlespacing

\maketitle

\vspace{-.2in}

\begin{abstract}
We introduce a non-parametric method to recover physical probability distributions of asset returns based on their European option prices and some other sparse parametric information. Thus the main problem is similar to the one considered foir instance in the Recovery Theorem by Ross (2015), except that here we consider a non-dynamical setting.
The recovery of the distribution is complete, instead of estimating merely a finite number of its parameters, such as implied volatility, skew or kurtosis. The technique is based on a reverse application of recently introduced Distribution Matching by the author and is related to the ideas in Distribution Pricing by Dybvig (1988) as well as comonotonicity.
\end{abstract}

\medskip


\ \\ 


\thispagestyle{empty}

\clearpage

\onehalfspacing

\section{Introduction}
\noindent In the seminal papers of Black and Scholes (1973) and Merton (1973) the European style option pricing model was introduced, later referred to as the Black-Scholes-Merton (BSM) model, which became the gold standard in equity options pricing. As any idealized model, it assumes several conditions which are not financially realistic. For instance, the model allows limitless borrowing, it assumes neither transaction costs nor liquidity concerns, and the stock price is assumed to follow a geometric Brownian motion. The last assumption can actually be viewed as a long list of different econometric assumptions with a varying degree of technicality. For instance, there are no discontinuous jumps, the random shocks are not autocorrelated, and the distribution of the stock price at any given moment is log-normal. 

Mandelbrot (1963) and Fama (1963) observed that financial asset returns are typically non-normal. The deviation from normality can be easily detected by estimating 
moments from returns data. 

Deviation of the Arrow-Debreu state price density from the log-normal shape, which is the case in the BSM model, has been documented as well. This is sometimes known as the \emph{volatility smile} due to the fact that the fitted volatility parameter are not constant with respect to strike prices or moneyness. Volatility surfaces including different maturities are applied daily by finance practitioners, and these functions are also under active theoretical study, recently by Carr and Wu (2009, 2016), Engel and Figlewski (2015) and discussed by Fengler (2005).

To reconcile the BSM model or some other structural 'benchmark' model with the observed statistical deviations in return, several authors have suggested 
different approaches belonging to different types\footnote{Kristensen and Mele (2011) use the term 'auxiliary' in place of 'benchmark' having a somewhat similar philosophy.}. These include structural approaches; for instance Heston (1993) and  Madan, Carr and Chang (1998) 
have developed flexibly parametrized models, which can be more closely calibrated to the empirical returns. The non-structural methods are typically more faithful to given empirical distributions. These methods include mixed log-normal distributions, which can further be seen as parametric; Edgeworth and Gram-Charlier expansions, Hermite polynomial approximations (semi-parametric) and then non-parametric methods including implied trees and maximum entropy principle. The parametric methods are studied for instance by Bahra (1996) and Figlewski (2010). The expansion methods are investigated for instance by Airoldi (2005), Jarrow and Rudd (1982), Corrado and Su (1996), Hermite approximation by Madan and Milne (1994) and implied trees by Rubinstein (1994) and Jackwerth (1999) and entropy in finance by Stutzer (1996).

Since there are some deviations both in the physical side of the pricing models, observed statistically from equity returns, and the risk-neutral side, observed from option prices,
it is natural to ask how these are related. This calls for a connection between risk-neutral and physical densities. Such a natural connection has many names, Stochastic Discount Factor (typically in Finance), Pricing Kernel or Radon-Nikodym Derivative (Quantitative Finance) and marginal utility of market portfolio of a representative investor (Economics). This is the ratio of the Arrow-Debreu state price density and the physical probability density at a given state, i.e. equity (index) value. Pricing kernel is a very central, frequently applied tool in finance, and its unexpected shape is studied by Bakshi and Chen (1997), Bakshi, Madan and Panayotov (2010) and Songa and Xiu (2016). The physical and risk-neutral characterisitics of the observed densities have been connected by Chernov and Ghysels (2000), Bakshi, Kapadia and Madan (2003) and Chalamandaris and 
Rompolis (2012).

Although the treatment may be holistic, analyzing both physical and risk-neutral densities simultaneously together with their interrelations brings up two natural (sub)questions.
Namely, how does the physical density of returns (or its deviations) affect the risk-neutral one, and vice versa?
Schl\"{o}gl (2013) and Xiu (2014) are able to represent a rather general distribution as an infinite polynomial expansion. The expansions can be associated to infinite portfolios of European style derivatives on the equity in the BSM model. The value of the portfolio then serves as the estimated value of the asset. There are techniques to choose expansions in such a way that given kurtosis and skew can be satisfied within limits. Then there is another question: How can physical or risk-neutral skew or kurtosis be extracted from option prices? This is a parametric problem, adressed by Corrado and Su (1996), Bakshi, Kapdia and Madan (2003). Another parametric problem involving risk-aversion of a representative agent implied by option prices in investigated by Bliss and Panigirtzoglou (2004). 

Due to changes in the pricing kernel over time, on one hand, and the large volumes of options being traded, on the other hand, it seems reasonable to depart from the strict version of pricing by dynamical arbitrage paradigm. One may argue that the option prices involve market sentiment on future returns of the underlying. For instance, the average volatility of an equity price index on a given future time interval is up to speculation while it is relevant in pricing options on the index.
Here we posit that option prices on an equity price index reflect ex-ante the view of the markets on the return of the index, 
up to some extent. We will device a method to construct \emph{implied physical distributions} of asset returns. In a sense this extends the notion of implied volatility $\widehat{\sigma}$ to a non-parametric setting because the same $\sigma$ 
parameter appears both in the physical and the risk-neutral distributions in the BSM model. 

To this end, we continue developing a novel pricing technique, termed as \emph{distribution matching} which is initiated in Talponen (2018). The general idea in this matching exhibits some similarity to orthogonal polynomial expansions or implied trees, in the sense that the return distribution of the asset being priced is replicated. Here the distribution-replication is performed by transforming the state space of a benchmark model. This idea is closely related to the distribution pricing by Dybvig (1988). The distribution of the asset return is not merely approximated but is matched exactly. The technique is non-structural and semi-parametric in nature. This means that in the first stage the benchmark, in this paper the BSM model, is calibrated according to market data (i.e. the parametric part). In the next stage the  transforming of the state space takes place and this procedure is fully non-parametric. 
 
Intuitively, here the asset analyzed is assumed to approximately follow the BSM model. However, the dynamics of asset prices are not considered here, but instead the assets are treated as single-step random cash flows. This is mostly sufficient, at least for the purposes of the relative pricing of assets. Here then the \emph{physical} distribution of the asset value is estimated from option prices. Our aim is similar to the Recovery Theorem by Ross (2015). Most of the research on implied distributions is concerned with the estimation of \emph{risk-neutral} distributions in particular. It should be noted that the risk-neutral dynamics may be far apart from the physical dynamics, as, e.g., in the case of the variance gamma process, studied by Madan, Carr and Chang (1998). Therefore, inference of physical distributions solely based on option prices in the general case would be impossible. However, the BSM model is more tractable, as the risk-neutral density of sample paths is a function of the terminal values. Here additionally sparse parametric information is used on the statistical returns of the priced asset and the benchmark security.\footnote{For example, Melick and Thomas (1997) argue that in their study the terminal distribution provides sufficient information for pricing, so that fixing the whole stochastic process is not required.}

In distribution matching there is an empirical physical density of asset returns which is coupled with a physical and risk-neutral densities of a benchmark pricing model of derivatives. The technique then produces a risk-neutral density 
estimate corresponding to the given empirical distribution. This procedure can be reversed and this is the crux in this paper. Namely, we run the following thought experiment: Assume that a representative investor applied distribution matching to an emprical distribution, together with BSM benchmark model with predictable parameters. Then European style options become priced based on the risk-neutral distribution. Now, suppose that the prices of the resulting options are known, what is the empirical distribution employed by the representative investor? 

A significant advantage of the technique here is that the complete estimated distribution with high resolution is readily obtained, instead of finitely many moments or parameters. Consequently, there is no burden of checking whether there actually exists a probability distribution with the given estimated moments and other parameters, cf. Longstaff (1995).

We will illustrate the recovery of physical densities of asset price returns by analyzing a given SP500 price index option chain. Thus, this recovery uses distribution matching in reverse. We also show how distribution matching can be applied in 
smoothing the state price density resulting from the noisy option price data.

\section{Pricing by distribution matching}\label{sect: recall}

In this paper the pricing technique introduced in Talponen (2018), Distribution Matching, is applied, chiefly in reverse, to estimate an ''implied physical distributions'' of a given asset. Let us first explain the direct version of the technique.

Pricing by distribution matching involves an asset being valued, $S_1$, and a benchmark security, $S_2$ which is traded
and has abundantly sorts of European style options written on it. The $1$-step model is focused on a given time interval $[0,T]$ and 
in particular the distributions of $S_1 (T)$ and $S_2 (T)$ which involve separate models, $\mathcal{M}_1$,  $\mathcal{M}_2$. The time $T$ is also the expiry of the options on $S_2$. Ideally, the asset being valued $S_1$ and the benchmark security $S_2$ are very highly correlated. Thus the setting is static here, complementary to different approaches exploiting the dynamics of the assets, see e.g. Adersen et al. (2015) and Ross (2015).

Next we will explin the assumptions, or rather the thought experiment behind distribution matching.  Hocquard et al. (2012) propose a method for tailoring funds accoridng to 
a target distribution with a similar philosophy, cf. Halperin and Itkin (2014). Our construction is continuous-state but its discrete version is closely related to 
Rubinstein's (1994) implied trees and the Marketed Asset Disclaimer in real options analysis.

In distribution matching one first considers portolios $\Pi_0$ of digital options on $S_2$ with expiry $T$ in such a way that the value of the portolio $\Pi_0(T)$, considered at time $t=0$, is highly correlated with $S_2 (T)$ and the distribution of the portfolio value at time $T$ is close to that of $S_1 (T)$, in symbols 
\[\mathrm{Corr} (\Pi_0 (T), S_2 (T))\approx 1, \quad \Pi_0 (T) \sim S_1 (T)\ \text{approximately}.\]
These portfolios are finite and as such the distributions are rough approximations of that of $S_1 (T)$, and
at this stage the portfolios are by no means unique.

As a response to the above issues one passes to the limit in the process of refining these portfolios in such a way that the asymptotic payoff distribution of the portfolio matches \emph{exactly} the distribution of $S_1 (T)$. 
This leads to an analysis of physical and state price distributions with a particular state space transformation.
The resulting idealized portfolio of infinitesimal Arrow-Debreu securities is not \emph{ad hoc} anymore at this stage; instead it is a unique arrangement of options such that some natural properties of the pricing functional are satisfied.
In an ideal case where the benchmark security and the asset are very highly correlated, the ideal portfolio $\Pi$, which can be seen as a European style derivative on $S_2$, satisfies
\begin{equation}\label{eq: fund}
\mathrm{Corr} (\Pi (T), S_1 (T))\approx 1, \quad \Pi (T) \sim S_1 (T).
\end{equation}

Note that $\Pi (T)$ then statistically hedges $S_1$ but the hedging notion is actually much stronger due to high 
correlation. If the correlation was perfect, then the portfolio or derivative would be a perfect static hedge for $S_1$ in our single-step model. 

In distribution matching the time $t=0$ price of the asset is modeled by the corresponding value of the portfolio:
\[\widehat{S_1 (T)} = \Pi (0).\]
Intuitively, the loss of information resulting from the imperfect correlation in \eqref{eq: fund} is partially patched up 
by the market information imported in the model via option prices.

The formula derived for $\Pi (0)$ in Talponen (2018) is as follows:
\begin{equation}\label{eq: value} 
\Pi (0)=\int_{a_1 }^\infty s \frac{\phi_{1}(s)}{\phi_{2}(\K(s))} q_2 (\K(s))\ ds\
\end{equation}
where the state $s$ stands for the numerical values of $S_1 (T)$ and $\phi_1$ and $\phi_2$ are the continuous physical distributions of $S_1 (T)$ and $S_2 (T)$, respectively, $q_2 $ is the state price density of $S_2 (T)$ and the state space transformation $\K$ maps $S_1 (T) \mapsto S_2 (T)$ between the models $\mathcal{M}_1$ and $\mathcal{M}_2$. It satisfies the differential equation
\[\K'(x)=\frac{\phi_{1}(x)}{\phi_{2}(\K(x))},\ \K(x_0 )=y_0 \]
which can also be easily used in solving the transform numerically. The initial state and the initial value, $x_0$ and $y_1$, 
are theoretically the essential minimal values of $S_1 (T)$ and $S_2 (T)$ (typically both $0$) but for numerical reasons we apply positive values in practice.

The state price density can be derived in principle by using the Breeden and Litzenberger (1978) representation, together with a smoothening kernel regression. These are explained here below.  

Above we explained how the portfolio $\Pi$ is constructed as a limit of finite portfolios. This is essentially static hedging, see e.g. Derman et al. (1995), Carr et al. (1998).
The portfolio $\Pi$ can alternatively be viewed as a European style derivative on $S_2$ with the following properties:
\begin{enumerate}
\item The derivative and $S_1 $ have the same value distribution at maturity $T$. 
\item The payoff of the derivative is an absolutely continuous strictly increasing function on $S_2 (T)$, i.e. $\Pi(T)$ and $S_2 (T)$ are comonotone. 
\end{enumerate} 
The latter condition typically implies that the derivative payoff and $S_2 (T)$ are highly correlated. In a nutshell:

\emph{In distribution matching we price an asset by constructing a European style derivative on a benchmark security such that the payoff distribution of the derivative matches that of the asset and the payoff is increasing and continuous.}

The relationship between the derivative's payoff function $f$ and the state space transform $\mathcal{K}$ is simple:
\[f=\mathcal{K}^{-1}  .\]

The distribution matching price formula \eqref{eq: value} yields a natural candidate for the unobserved SPD on $S_1 (T)$:
\begin{equation}\label{eq: estQ}
\widehat{q}^{\mathrm{DM}}_1 (x) = \frac{\phi_{1}(x)}{\phi_{2}(\K(x))} q_2 (\K(x)).
\end{equation}
Let us collect the characteristics of the speed of the transform:
\begin{equation}\label{eq: speed}
\frac{\widehat{q}^{\mathrm{DM}}_1 (x) }{q_{2} (\mathcal{K}(x))}=\frac{\phi_{1} (x)}{\phi_{2} (\mathcal{K}(x))}= \mathcal{K}'(x) .
\end{equation}

This states that $\mathcal{K}$ preserves \emph{both} the risk-neutral probabilities and the physical probabilities of terminal value
intervals between the models $\mathcal{M}_1$ and $\mathcal{M}_2$:
\[\mathbb{P}_1 (S_1 (T) \leq x)=\mathbb{P}_2 (S_2 (T) \leq \mathcal{K}(x)),\]
\[\mathbb{Q}_1 (S_1 (T) \leq x)=\mathbb{Q}_2 (S_2 (T) \leq \mathcal{K}(x)).\]

\section{Estimation of ex-ante physical densities from option prices}\label{sect: estPhys}

This section involves the main contribution of the paper. We are able to provide a reasonable estimate for the physical distribution of an asset price based on options written on the asset. It is noteworthy that an ex-ante \emph{physical} distribution is in fact estimated here, although in the literature the risk-neutral ones have been the object of extensive study. Also, a full distribution is semi-parametrically estimated, not merely some moments or parameters of it.

Intuitively, the underlying assumption here is that the asset somewhat closely follows the BSM model. Here $[t,T]$ is the time interval of interest, $S_t$ is the current value of asset and we will analyze the physical and risk-neutral distributions of $S_T$. 

The reasoning behind this technique is the following. Suppose that a representative investor in the economy has an ex-ante
belief about the physical density of the rate of return of the asset in question over the given time step. She has access to 
predictable financial measures $\widehat{r}$, and $\widehat{\mu}$ and $\widehat{\sigma}$ involving the asset,
which she applies as the the BSM parameters. The resulting BSM model then serves as a benchmark in the process of pricing the asset. This appears particularly fruitful in the case there exists a liquid 'benchmark' security in the economy with European style options on it such that the security and option prices follow exactly the calibrated BSM model and the security is almost perfectly correlated with the asset analyzed. Suppose that the representative investor then applies Distribution Matching to model the SPD of the asset in question. If we know the resulting option prices and can predict the BSM parameters applied 
by the representative investor, then we can recover her ex-ante physical distribution of the asset's prices.

The resulting implied physical distribution can be regarded as a non-parametric analogue of implied volatility.
We are not claiming, a priori, that the return distributions estimated ex-post should perfectly reflect the ex-ante ones, 
even in the long run on some average sense. The ex-ante distributions could exhibit for instance Knightian uncertainty, and related ambiguity aversion, or simply systematically biased predictions. The problematic relationship between implied volatility and realized volatility is discussed by 

The implied volatility from stock options can explain the realized volatility better than can the historical volatility (Szakmary et al. 2003).

Brous et al. (2010), see also Chalamandris and Rompolis (2012), 
Carr and Wu (2016).

\subsubsection{Method inputs}
We will apply the following market data and steps regarding an asset and European style calls on it:
\begin{enumerate}
\item The current price of the asset, $S_t$.

\item We analyze options on $S_T$ to estimate the state price density $\widehat{q}_1$ on the asset, based on the estimation technique explained subsequently.

\item We form for BSM model purposes a short rate estimate $\widehat{r}$. This is obtained from the yield 
of an investment grade zero-coupon bond with roughly matching maturity.

\item The implied volatility $\widehat{\sigma}$ is estimated by using $\widehat{r}$ and $\widehat{q}$ such that the corresponding BSM model risk-neutral density, 
depending on $\widehat{r}$ and $\widehat{\sigma}$, has the same interquartile range as the risk-neutral density corresponding to $\widehat{q}_1$. 

\item An estimate for a BSM model trend $\widehat{\mu}$ is ideally forecasted by a specialist. 
Here we form the estimate $\widehat{\mu}$ according to recent realized returns (annualized BSM parameters) :
\[\widehat{\mu} = \ln \frac{S_t}{S_{t_0}}\]
according to the BSM model expected return formula $\mathbb{E}_{\mathbb{P}} S_T = S_{t} e^{\mu (T- t)}$. 

\item The benchmark security is assumed to follow the BSM model.
Based on the estimated parameters $S_t$, $\widehat{r}$, $\widehat{\sigma}$ and $\widehat{\mu}$,
we obtain a BSM model physical density $\widehat{\phi}_2$ and the state price density 
$\widehat{q}_2$ on the benchmark security. 
\end{enumerate}

The location parameter of the physical BSM density is underdetermined here, and, consequently, the location of both the densities will be that as well.
However, one is interested in the general shape of the distribution pertinent to skew and kurtosis and therefore the issue with the location is not crucial.

\subsubsection{Solving implied physical density}

Next the physical density estimate is given.
The estimation for the distribution of the asset is then obtained by applying \eqref{eq: estQ} 
\[\widehat{q}^{\mathrm{DM}}_1 (x) = \frac{\phi_{1}(x)}{\phi_{2}(\K(x))} q_2 (\K(x))\]
in reverse as follows:
\begin{equation}\label{eq: estphi}
\widehat{\phi}^{\mathrm{IPD}}_{1}(x)  = \frac{\widehat{q}_1 (x) }{\widehat{q}_2 (\K(x))} \widehat{\phi}_{2}(\K(x)),
\end{equation}
where $\mathcal{K}$ satisfes \eqref{eq: speed}, thus 
\[\frac{\widehat{q}_1 (x) }{\widehat{q}_{2} (\mathcal{K}(x))}= \mathcal{K}'(x) . \]

The transformation $\mathcal{K}$ is straight-forward to solve for any initial condition $\mathcal{K}(x_0)=y_0$ numerically by Euler's method: 
\ \\

\vbox{
\begin{framed}
Let $x_1 = x_0$ and $y_1 =y_0$; $\% \ x_i = x$, $y_i = \mathcal{K}(x)$\\

For i=1 to N;\\

\quad Let $\widehat{\phi}^{\mathrm{IPD}}_1 (x_{i})=\frac{\widehat{q}_{1}(x_i )}{\widehat{q}_{2}(y_i )} \widehat{\phi}_2 (y_i )$; \\

\quad Let $x_{i+1}=x_i +h$ and $y_{i+1}=y_{i}+ \frac{\widehat{q}_1 (x_i ) }{\widehat{q}_{2} (y_i )} h $;\\

Next;\\

\end{framed}
}

The initial values $x_0$ and $y_0$ should be the $p$-quantiles for the same $p$ in the respective measures,
$\mathbb{Q}_1$ and $\mathbb{Q}_2$, since $\mathcal{K}$ models a measure-preserving transform. From the numerical point of view, the failure of this results in 
trivial asumptotics of $\mathcal{K}' (s)$ as $s\to \infty$, either a rapid convergence to $0$ or a divergence to $\infty$.

Below we have the explicit expressions for \eqref{eq: estphi} employing to the BSM densities:
\[\widehat{\phi}^{\mathrm{IPD}}_{1}(x)  = \widehat{q}_1 (x)  \exp \left[-\frac{\Delta }{2\widehat{\sigma}^2 (T-t)}\right],\]
\[\Delta=(\ln(\mathcal{K}(x) /S_t) - (\widehat{\mu}-\widehat{\sigma}^2 /2)(T-t))^2 -(\ln(\mathcal{K}(x) /S_t) - (\widehat{r}-\widehat{\sigma}^2 /2)(T-t))^2,\]
\[\mathcal{K}'(x) = \widehat{\sigma}\sqrt{2\pi(T-t)}\widehat{q}_1 (x) \mathcal{K}(x) \exp \left[{\frac{(\ln(\mathcal{K}(x) /S_t) - (\widehat{r}-\widehat{\sigma}^2 /2)(T-t))^2 }{2\widehat{\sigma}^2 (T-t)}}\right] .\]

Thus, for instance, we may estimate (physical) skew and kurtosis of index returns implied by option prices.
Similar philosophy appears in Bakshi, Kapadia and Madan (2003). The benefit here is that from the implied physical 
distribution we effortlessly obtain any moment, or in fact the value of any functional in our diagnostics kit, for instance a risk measure.  

\subsection{Illustration with a SP500 option chain}
Next we will study a given, rather generic recent SP500 price index (SPX) option chain\footnote{The investigated option chain data, bid prices of SPX1816C1000-E  - SPX1816C3000-E, was taken from CBOE web pages at Sep 19, 2017 @ 08:54 ET with 178 days until expiry. The last SPX trade was quoted at the price 2503.87.}. We will numerically illustrate the implied distribution recovery technique. This yields a rather high resolution estimate for implied physical return distribution.

The empirical risk-neutral distributions implied by the option data are very rough in practice. 
At this stage we take the smooth SPD estimate as given.  We will show later in this paper how distribution matching can be applied to smoothen the implied risk-neutral distribution.

We rely on the benchmark BSM model both in distribution matching and in estimating the parameters from empirical data.
The short rate $r$ is estimated from the annual yield quote of T-bills with maturity 3/15/2018 at the given date. Thus 
$r=\log(1.0116)=0.012$. Applying this rate we will find such an implied volatility $\widehat{\sigma}$ such that the RN distribution of the BSM model, thus with parameters, $S_0 =2503.87$, $T=178/365$, $r=0.012$ and $\widehat{\sigma}$ has the same IQR as the SPD estimated from the option chain. Thus we obtain $\widehat{\sigma}=0.0982$; for comparison
the implied volatility at-the-money is $\widehat{\sigma}_{IV}=0.1028$. Note the exceptionally low implied volatility.

{
\centering
\begin{figure}[H]
     \centering
     \subfloat{\includegraphics[width=0.5\linewidth]{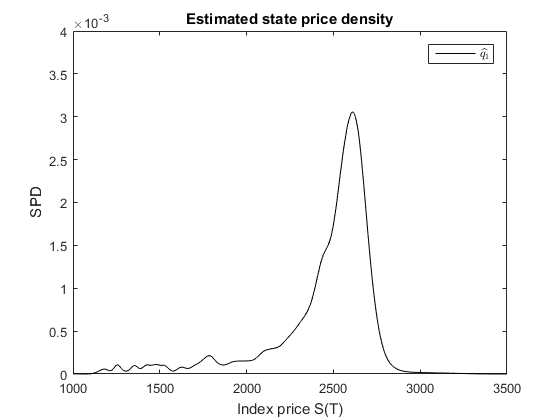}}
     \subfloat{\includegraphics[width=0.5\linewidth]{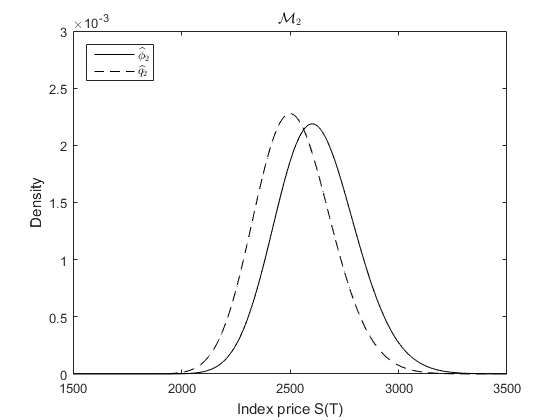}}
     \caption{The implied physical density estimation of ${\widehat{\phi}}^{\mathrm{IPD}}_{1}$ requires $3$ distributions and the transformation $\mathcal{K}$. The model $\mathcal{M}_1$ risk-neutral density $\widehat{q}_1$ is on the left. 
     The $\mathcal{M}_2$ physical and risk-neutral densities $\widehat{\phi}_{2}$ and $\widehat{q}_2$ appear on the right.}

\end{figure}
}

The trend parameter $\mu$ of the BSM benchmark model is required. The intended end user of the technique may 
consult a specialist in forecasting $\mu$. Here we will base our choice on the $9.84\%$ price index return in 2016 and we obtain $\mu=\log(1.0984) = 0.094$.

We will propose the methodology required for estimating the SPD subsequently. 
The price quotes in the option chain become sparse far away from at-the-money and thus the quality of the SPD data 
is not homogenous. Therefore we will apply a large initial value $x_0=1300$ in the ODE which yields a  
numerical solution to \eqref{eq: estphi}. Since in our model $\mathcal{K}$ is $\mathbb{Q}_1$-$\mathbb{Q}_2$-measure-preserving, 
the image $\mathcal{K}(x_0 )=2219.18$ and $x_0$ are chosen in such a way they become the same quantiles in their respective distributions, 
$\mathbb{Q}_1$ and $\mathbb{Q}_2$.\footnote{In Distribution Matching $\mathcal{K}$  is constructed to be $\mathbb{P}_1$-$\mathbb{P}_2$-measure-preserving, 
i.e. preserving quantiles. Then $\mathbb{Q}_1$ is defined so that $\mathcal{K}$ becomes $\mathbb{Q}_1$-$\mathbb{Q}_2$-measure-preserving.}  
 
{
\begin{figure}[H]
\centering
{\includegraphics[width=\textwidth]{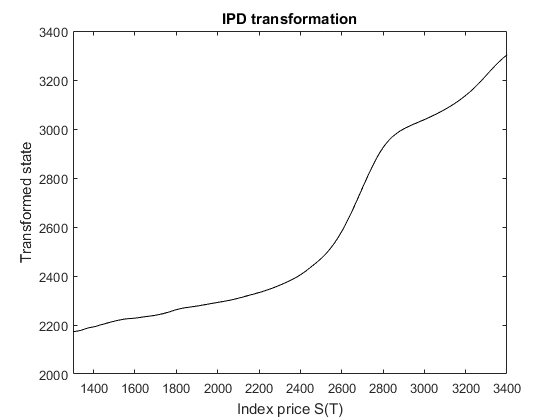}}
\caption{ }
\label{Fig: K}

\end{figure}
}

Here is the transform $\mathcal{K}\colon S_1 (T) \mapsto S_2 (T)$ applied.
If the portfolio constructed in Distribution Matching is interpreted as a European style derivative on the security in 
$\mathcal{M}_2$, then $\mathcal{K}^{-1}$ is its payoff function. Thus the above Figure \ref{Fig: K} depicts the inverse mapping of a suitable derivative payoff profile.

The above transform $\mathcal{K}$ is solved numerically from the SPD $\widehat{q}_1$, estimated from the option chain, 
the BSM benchmark model ($\widehat{\phi}_2$ and $\widehat{q}_2$, depending on $S_0$, $T$, $\widehat{r}$, $\widehat{\sigma}$ and $\widehat{\mu}$), 
and \eqref{eq: speed}. Then using $\mathcal{K}$ in \eqref{eq: estphi} produces an estimate $\widehat{\phi}_{1}^{\mathrm{DM}}$.

{
\begin{figure}[H]
\centering
\includegraphics[width=\linewidth]{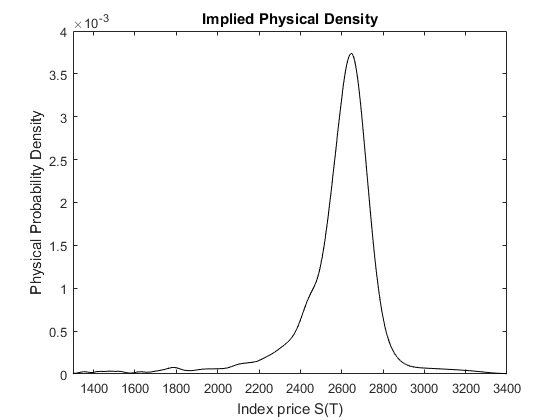}
\caption{ }
\end{figure}

}
This example clearly deviates from log-normality, it exhibits large kurtosis (11.26) and negative skew (-1.836) in a fairly typical manner, cf. Conrad et al. (2013).

\subsection{The soundness of the IPD estimation}

The basic properties of the distribution mathcing are treated in Talponen (2018) and they essentially imply the following result which states that the IPD technique correctly estimates the physical density of a BSM model $\mathcal{M}_1$ provided that some natural assumptions are met. An initial condition $\mathcal{K}(0)=0$ can be valid below if 
we accept weaker solutions to the ODEs.

\begin{proposition}\label{prop: coincide}
Let $\mathcal{M}_1$ and $\mathcal{M}_2$ correspond to BSM models whose parameters satisfy that the
market prices of risk coincide:
\[\lambda_1 := \frac{\mu_1 - r_1}{\sigma_1} = \frac{\mu_2 - r_2}{\sigma_2} =: \lambda_2 .\]
Suppose that we form an IPD estimate $\widehat{\phi}_{1}^{\mathrm{IPD}}$ by using the risk-neutral 
densities $q_1$, $q_2$ and the physical density $\phi_2$ as follows. Choosing positive initial values $x_0$ and $\mathcal{K}(x_0)$ such that 
\[\mathbb{Q}_1 (S_1 (T) \leq x_0)=\mathbb{Q}_2 (S_2 (T) \leq \mathcal{K}(x_0))\]
and considering \eqref{eq: speed} and \eqref{eq: estphi} yields a unique solution $\widehat{\phi}_{1}^{\mathrm{IPD}} (x)$, $x\geq x_0$. Then 
\[\widehat{\phi}_{1}^{\mathrm{IPD}} (x) = \phi_1 (x),\quad x\geq x_0 .\]
\end{proposition}

More generally, if the models correspond to suitably 'isomorphic' stock evolutions with a common driving Ito process 
$X_t$ and the local market prices of risk coincide, $\lambda_1 (t) = \lambda_2 (t)$, then we have a similar
conclusion as above. See Talponen (2018) for the details and a proof.

\subsection{The choice of the trend parameter}
The weak point in our recovery technique is that the choice of $\mu$ apllied in here the benchmark model was rather ad hoc. In fact, in the BSM model the risk-neutral density does not depend on $\mu$ and therefore the trend cannot be deduced form option prices. 
Thus the estimation of the implied physical density of returns involves a subproblem involving the trend. Our estimation technique does not provide a clear answer to the latter question and it must be forecasted separately. The interpretation of $\mu$ becomes quite subtle here, it is the time-$t$ expected rate of return over the interval $[t,T]$
according to the belief of the representative investor (or the general sentiment of the markets).

The choice of the trend does affect the estimated physical distribution. To illustrate this we used different $\mu$ parameter values in the recovery and the measures of the resulting implied physical distributions are listed below.

\begin{table}[h!]
\label{table:1}
\begin{center}
\begin{tabular}{|l |c c c c|}\hline
\backslashbox{$\mathrm{Measure}$}{$\mu$} & 0.000 & 0.050 & 0.094 & 0.015 \\
\hline
$\mathrm{Mean}$ & 2576.5 & 2585.7 & 2593.8  & 2606.4\\
\hline
$\mathrm{Median}$ & 2614.5 & 2621 & 2626.5 & 2634.7\\
\hline
$\mathrm{STD}$ & 205.3148 & 206.3828 & 207.7087 & 208.8049\\
\hline
$\mathrm{Skew}$ & -2.0225 & -1.9329 & -1.8358 & -1.6807 \\
\hline
$\mathrm{Kurtosis}$ & 11.3727 & 11.3379 & 11.2558 & 11.1316\\
\hline
\end{tabular}
\end{center}
\caption{The dependence of statistical measures of the IPD on the trend parameter $\mu$ applied in the estimation.}
\end{table}

\bigskip
It can be anticipated that the choice of $\mu$ affects the mean and the median of the implied physical distribution but it even appears to have a considerable effect on the skew.

This appears to be mainly an issue with the numerics. Since the choice of $\mu$ affects distribution function of the benchmark physical distribution, say, $\mathcal{K}(1300)$ become $p$-quantiles for different $p$-values in different models. Recall that in this framework $\mathcal{K}$ is measure-preserving both for physical and risk-neutral measures.  

To reconcile with these features we experimented adjusting the transform in such a way that in different models, corresponding to different $\mu$, the \emph{physical} probabilities $\mathbb{P}_2 (S_2 (T)<\mathcal{K}(1300))$, which depend on $\mu$, are normalized to have the same value. This is performed by choosing suitable initial values $\mathcal{K}(x_0 )=y_0$ for $x_0 =1300$ in the ODE \eqref{eq: speed}. 

{
\centering
\begin{figure}[H]
     \centering
     \subfloat{\includegraphics[width=0.5\linewidth]{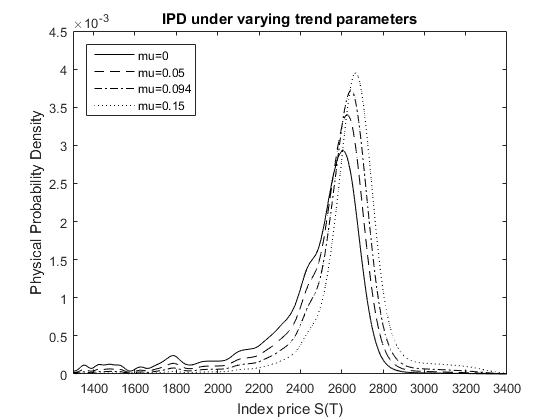}\label{figure1}}
     \subfloat{\includegraphics[width=0.5\linewidth]{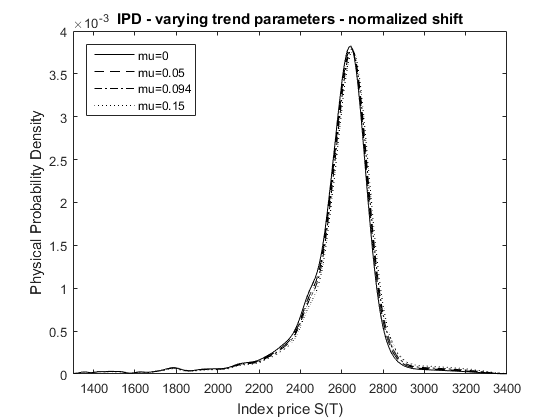}\label{figure2}}
     \caption{The implied physical distribution before and after normalizing the quantiles. This means that for each $\mu$ the value $\mathcal{K}(1300)=y^{(\mu)}_0$ is adjusted to be the same $\mathbb{P}_2$-quantile.}

\end{figure}
}
\noindent This suggests that even if the trend cannot be forecasted, the general shape of the distribution is rather stable under varying $\mu$. The IPDs are very similar
up to a kind of shift operation.

These findings should be compared to Proposition \ref{prop: coincide}. Note, however, that here the market prices of risk were not normalized. 

\section{Semiparametric estimation of state-price densities}\label{sect: SPD}
Breeden and Litzenberger (1978) show that the SPD can be obtained in a model-free way from European call prices by the following elegant formula
\[q(x)=\frac{\partial^2 C(S_t , t,T,K)}{(\partial K)^2}\bigg\vert_{K=x}.\]
This can easily be turned into a numerical method which approximates the SPD by means of numerical differentiation. There are some issues with such a method, the obtained SPD is typically very rough and exhibits some negative values which in principle correspond to arbitrage opportunities. 

We estimated the SPD from the investigated SPX option chain by 2 different methods. Below on the left we applied the straight-forward numerical Breeden-Litzenberger approach restricting to positive values
and rescaling according to the benchmark bond price. 
 
{
\centering
\begin{figure}[H]
     \subfloat{\includegraphics[width=0.5\linewidth]{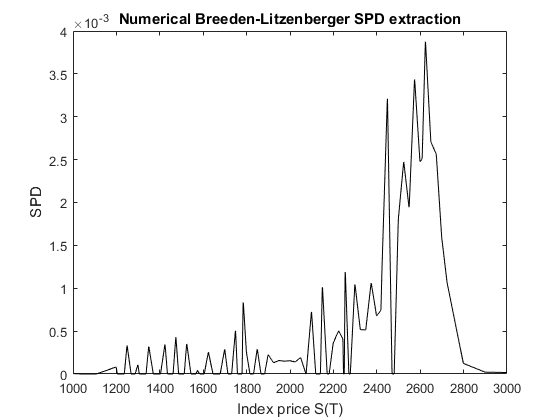}}
     \subfloat{\includegraphics[width=0.5\linewidth]{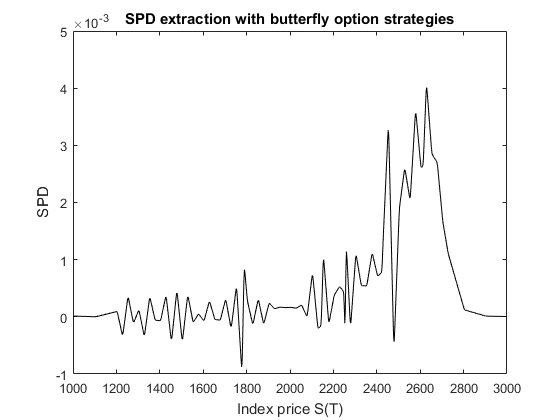}}
    \caption{Two different but related numerical SPD estimates.}
    \label{Fig: raw}
\end{figure}

}

For $\vartriangle\!\! K = (K_{i+1}-K_i )=1$ the numerical Breeden-Litzenberger SPD estimate has a simple formula
\begin{equation}\label{eq: widehat1}
\reallywidehat{q\left(K_{i+1}\right)} = C(K_i)-2C(K_{i+1})+C(K_{i+2}).
\end{equation}
This arises as a second order derivative calculated numerically with $h=\vartriangle\!\! K =1$.

To check the robustness of the computations, we also used 'iron butterfly' option strategies to estimate the \emph{slope} of the state price density at any given state. By concatenating a short and a long butterfly we obtain a payoff which is a pulse as follows:

{
\begin{figure}[H]
\centering
\includegraphics[width=5cm]{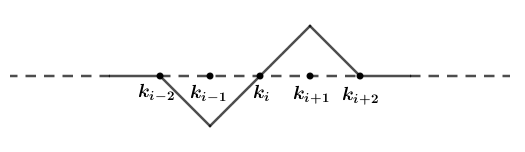}
\caption{The payoff profile of a concatenated short and long butterly.}
\end{figure}

By simple geometric resoning this derivative captures the average growth of the state price density on the interval $[K_{i}, K_{i+4}]$.
It can be built easily from long/short posititions call options and is priced accordingly\footnote{This relationship would be accurate if $q$ were linear on $[K_{i}, K_{i+4}]$ and holds approximately if $q$ is continuously differentiable and the interval is short.}:
\begin{equation}\label{eq: widehat2}
\reallywidehat{(q(K_{i+1})-q(K_{i-1}))} = -C(K_{i-2})+2C(K_{i-1})-2C(K_{i+1})+C(K_{i+2}).
\end{equation}

From the slopes we then formed a system of linear equations, imposing vanishing state price densities at the ends, $q(1000)=q(3000)=0$. Solving this exactly identified system of equations yields the right hand SPD in Figure \ref{Fig: raw}, where, for the sake of illustration, we made no effort to correct the negative values. Note that if the latter technique is also augmented similarly with positive restriction and scaling, then the end results of these methods become very similar. 
These two methodologies are closely related, it is instructive to observe that 
\eqref{eq: widehat1} and \eqref{eq: widehat2} yield
\[\reallywidehat{(q(K_{i+1})-q(K_{i-1}))} =  \reallywidehat{q\left(K_{i+1}\right)} - \reallywidehat{q\left(K_{i-1}\right)}.\]

The numerical Breeden-Litzenberger approach is widely applied and we will use this approach as a starting point due to its familiarity. However, at the end we propose an alternative method which solves some issues here.

\subsection{Non-parametric state price density estimation by kernel smoothing of a state space transformation}

Above we recalled the numerical Breeden Litzenberger technique which producing a 'raw' SPD candidate
from European style call option prices. The resulting noisy densities (see Figure \ref{Fig: raw}) may be inconvenient in view of numerical techniques and may not provide a theoretically satisfying model. 

There are several ways of smoothing state price densities, see Bondarenko (2003) for a convolution smoothing method
of RND and discussion on different approaches. In order for the method to be suitable for financial econometrics, the estimation should at least preserve arbitrage freeness. A\"{\i}t-Sahalia and Lo (1998) estimated state price densities from option price data by non-parametric kernel regression involving of the BSM model implied volatility (IV) and then they compute the SPD from the BSM model by using pointwise the estimated IV, cf. A\"{\i}t-Sahalia and Duarte (2003), Brunner and Hafner (2003). Thus they are in a sense leaning on a benchmark BSM model which we also do here. 
However, we will not deal with the IV, instead, we apply kernel smoothing on the state space transformation with a Gaussian kernel. Then we recalculate the SPD based on the benchmark BSM model SPD in tandem with the smoothened transformation. The resulting SPD will be smooth.

Although this approach is likely unanticipated, it perfectly follows the gist of density matching and appears justified when we specifically wish to keep a track of deviations around a particular bechmark model, such as the BSM model.

The main benefits of this approach are the following: Firstly, there is no burden in checking whether the resulting option price system is in fact arbitrage free (see Proposition \ref{prop: smooth}). Indeed, this readily follows from the simple fact that a convolution smoothened version of a non-decreasing sufficiently increasing function is smooth and strictly increasing. Secondly, the benchmark model applied may differ from the BSM one and may have several parameters.

Suppose that we have obtained an estimate for state price density $\widehat{q}$ as in the previous section. 
This is a non-negative function supported on positive numbers.
We present the smoothing scheme formally which must then be numerically implemented. 

\begin{enumerate}
\item Consider the SPD $\widehat{q}$ estimated (crudely) the from option prices.
We form a discount factor $D$ corresponding to the price of a zero-coupon bond according to $\widehat{q}$
\[D = \int \widehat{q} .\]
We apply this factor to convert SPDs to risk-neutral densities.

\item We fit a logNormal distribution $\psi_{BSM}$ (interpreted as a BSM model state price density) to $\frac 1 D \   \widehat{q}$. by choosing the parameters of the logNormal distribution in such a way that the medians and the IQRs of the two distributions coincide.

\item Then we define a continuous, increasing and measure-preserving transformation $\mathcal{K}\colon (0,\infty) \to (0,\infty)$, thus
\[\int_{0}^x   \widehat{q}(s)\d s =  \int_{0}^{\mathcal{K}(x)} D\ \psi_{BSM}(s)\d s,\quad x>0.  \]

\item Perform convolution smoothing on $\mathcal{K}$ in log-scale with smooth convolution kernel 
$\varphi_\varepsilon$ to obtain a smooth increasing function $\widetilde{\mathcal{K}}\colon (0,\infty) \to (0,\infty)$:
\[\widetilde{\mathcal{K}}(x) := [\varphi_\varepsilon \ast \mathcal{K} (\exp(\cdot))] (\ln x),\quad x>0 .\] 

\item Then we define $\widetilde{q}$ according to \eqref{eq: estQ} and \eqref{eq: speed} using the smoothened 
transform $\widetilde{\mathcal{K}}$ as follows:
\[\widetilde{q} (s) := \widetilde{\mathcal{K}}' (s)  D\ \psi_{BSM}(\widetilde{\mathcal{K}}(s)).\]

\end{enumerate}

Here we used the Gaussian kernel $\varphi_\varepsilon (x) = c_\varepsilon \exp(-x^2 / \varepsilon)$ in convolution smoothing.

\begin{proposition}\label{prop: smooth}
For a Gaussian convolution kernel $\varphi_\varepsilon$ we have

\begin{enumerate}

\item $\widetilde{q}$ is a strictly positive smooth function with 
\[\int \widetilde{q}(s)\d s = \int \widehat{q}(s)\d s =D .\]

\item $\widetilde{q} \to \widehat{q} $ in distribution as $\varepsilon \to 0$. 
Equivalently, the $\widetilde{q}$-prices of European style digital options converge to the respective $\widehat{q}$-prices
as $\varepsilon \to 0$. 

\end{enumerate}

\end{proposition}

{
\centering
\begin{figure}[H]
    \subfloat{\includegraphics[width=0.5\linewidth]{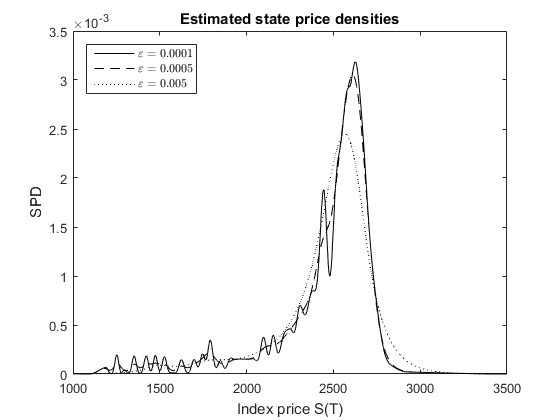}}
     \subfloat{\includegraphics[width=0.5\linewidth]{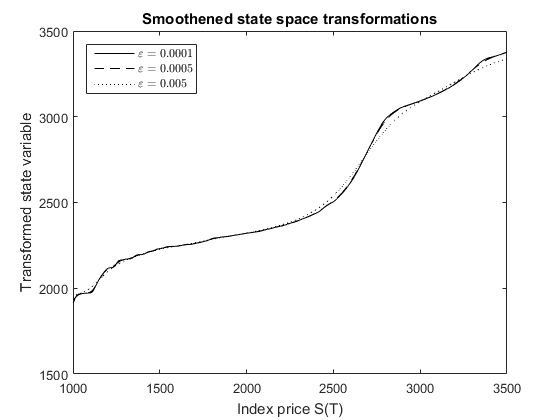}}
    \caption{Examples of SPD estimates with undersmoothened ($\varepsilon=0.0001$), reasonable ($\varepsilon=0.0005$) and oversmoothened ($\varepsilon=0.005$) kernel regression of $\mathcal{K}$.}
\end{figure}

}

Jackwerth and Rubinstein (1996) discuss the estimation of SPDs with smoothing and their findings are similar in shape to the above estimates.
The choice of the strength of smoothing is an art, may it be the bandwidth $h$ in Nadaraya-Watson kernel regression, or the $\varepsilon$ parameter in Gaussian convolution smoothing. 
Ideally the smoothened SPD should be both mildly oscillating and faithful to the given data. 
Above we excluded the distribution with $\varepsilon=0.0001$ because it is multimodal and excluded the case $\varepsilon=0.005$ because it involves too small a maximum likelihood compared to the non-smoothened SPD estimates.

One reasonable repeatable criterion in choosing the smoothing method parameter is using cross-validation, see e.g. Song and Xiu (2016). We applied parmaeter value $\varepsilon=0.0005$ in modeling the SPD used in the IPD construction.
Our rationale in choosing it was that it is (approximately) the smallest parameter resulting in essentially unimodal distribution.

{
\begin{figure}[H]
\centering
\includegraphics[width=11cm]{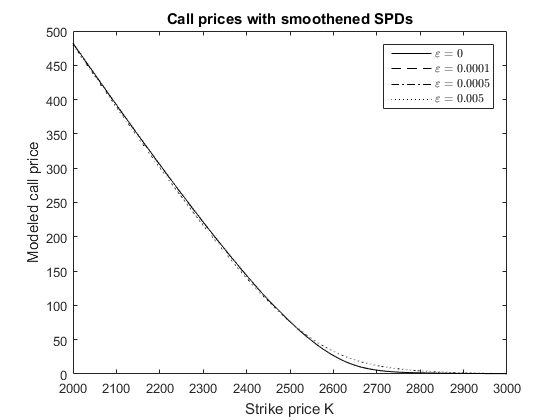}
\caption{The prices of calls are computed based on smoothened SPD after the numerical Breeden-Litzenberger extraction.}
\end{figure}

\begin{figure}[H]
\centering
\includegraphics[width=11cm]{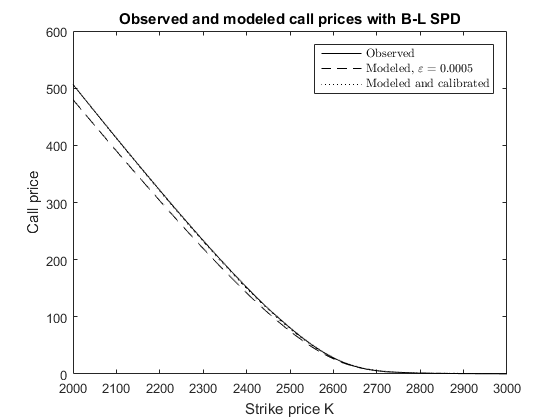}
\caption{The prices of calls are computed based on smoothened SPD after the numerical Breeden-Litzenberger extraction.
A calibration factor was solved at $K=2000$ and multiplying all modeled prices with this constant makes them virtually indistinguishable from the observed ones. }
\end{figure}

}

One may modify the above kernel smoothing technique to a local kernel regression to allow varying 
maturities, similarly as A\"{\i}t-Sahalia and Lo (1998). 

\section{SPD from option prices by quadratic programming}
Both the techniques applied here in extracting the raw SPD from the option data exhibit similar issues. Namely, the SPD estimates are highly oscilating and have negative values
which in principle translates to arbitrage opportunities. Such opportunities are usually not allowed in a SPD model and therefore we applied an ad hoc correction to get rid of 
the negative SPD values.

Recall the following integral equation, involving the SPD $q$, for the valuation of European style derivatives 
\[C_f (t) = \int f(s)\ q(s)\ ds\]
where the states $s=S(T)$ are the terminal values of the underlying security and $f(\cdot)$ is the payoff profile. In particular the European call options can be priced by 
\[C(t, K) =  \int \max(s-K,0)\ q(s)\ ds.\] 
We may discretize this as follows
\[C(t, K) \approx  \sum_i \max(s_j -K,0)\ q(s_j )\] 
where the goodness of the approximation depends on the continuity properties of $q$ and the mesh of $s_i$:s.
This leads to considering following linear equation system
\[\mathbf{c}=\mathbf{P}\mathbf{q}\]
where $\mathbf{c} = (C(t, K_i ))_i$, $\mathbf{P}=[\max(s_j -K_i ,0)]_{i,j}$ and $\mathbf{q}=(q(s_j ))_j$.
We are not claiming it has necessarily any non-negative solutions $\mathbf{q}$ for a given $\mathbf{c}$, even if $\mathbf{P}$ is a square matrix. However, we may find a unique non-negative vector $\mathbf{q}$ which is in a sense closest to a solution. This can be formulated as follows
\[\mathrm{argmin}_{\mathbf{q}\geq \mathbf{0}}\ (\mathbf{c} -\mathbf{P}\mathbf{q})^{T} W (\mathbf{c} -\mathbf{P}\mathbf{q})\]
where $W$ is $j\times j$ weight matrix. In the case with $W=I_{j\times j}$ the objective function becomes the square of the Euclidean norm $\|\mathbf{c} -\mathbf{P}\mathbf{q} \|^2$.
The solution is found by quadratic programming\footnote{We applied the lsqlin function in Matlab.}. The solutions for the 
SPD vector will be non-negative by definition and the quadratic nature\footnote{Recall that the $L^2$ norm is both the most uniformly smooth and uniformly convex norm among all norms.} of the optimization problem appears to smoothen them. This appears a plausible alternative to the numerical Breeden-Litzenberger approach. 

\subsection{Analyzing featured SPDs}
The above quadratic programming procedure produces a rather smooth, yet highly featured SPD which we alternatively used in the construction of an implied physical distribution of the returns. At this point we left the modeled SPD and the IPD undersmoothened by purpose in order to graphically illustrate the features of these distributions.

{
\begin{figure}[H]
\centering
\includegraphics[width=\linewidth]{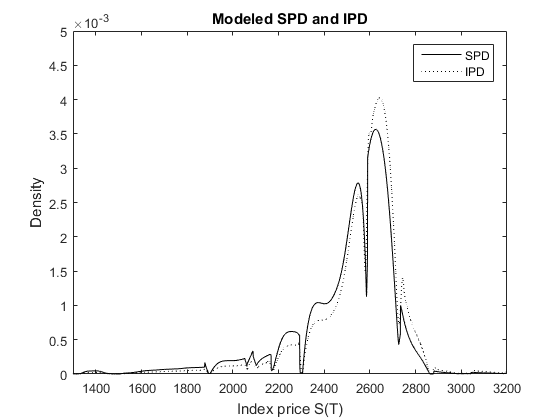}
\caption{Implied physical density based on a non-smoothened crude state-price density obtained by means of quadratic programming using identity matrix as the weight matrix.}
\end{figure}
}

We then formed an emprical distribution of the SPX rate of return corresponding to the time interval of our option chain.
The sampling was according to this time interval annually from fall 1964 to spring 2017. The implied volatility in our option chain was historically low and there was a bull market. We performed a heuristic state transform in an attempt to 
make the IPD better comparable to the empirical return distribution by 
increasing the standard deviation and decreasing the median of the IPD of the rate of return.

{
\begin{figure}[H]
\centering
\includegraphics[width=\linewidth]{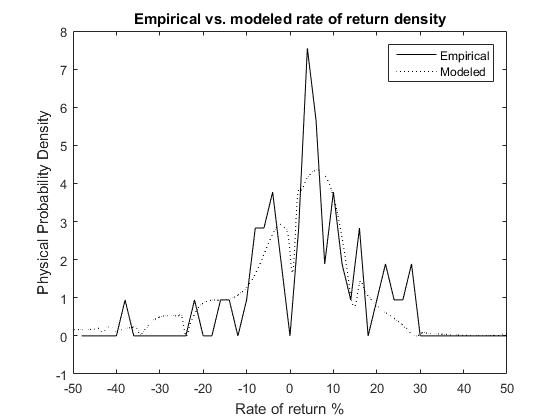}
\caption{Comparison of the modeled and empirical rate of return distribution. This was sampled for the similar time interval annually 1964-2017. An effort was made to shift and scale the IPD to make the distributions comparable.}\label{fig: emp}
\end{figure}
}

\bibliographystyle{chicago}

\begin{thebibliography}{}


\bibitem{AFT} T. G. Andersen, N. Fusari \and V. Todorov (2015). ''Parametric inference and dynamic state recovery from option panels'' Econometrica 83, 1081--1145.

\bibitem{airoldi}
M. Airoldi (2005). ''A moment expansion approach to option pricing'' Quant. Fin. 5, 89--104.

\bibitem{A\"{\i}t1} Y. A\"{\i}t-Sahalia, A. Lo (1998) ''Nonparametric estimation of state-price densities implicit in financial 
asset prices'' The Journal of Finance 53, 499--547.
  
\bibitem{A\"{\i}t1b } Y. A\"{\i}t-Sahalia, Lo, A., 2000. Nonparametric risk management and implied risk aversion. 
Journal of Econometrics 94, 9--51.  
  
\bibitem{A\"{\i}t2} Y. A\"{\i}t-Sahalia, J. Duarte (2003). ''Nonparametric option pricing under shape restrictions'' 
J. Econometrics 116, 9--47.

\bibitem{Bahra} B. Bahra (1996). Probability distributions of future asset prices implied by option prices.
Bank of England Quarterly Bulletin 36, 299--311.

\bibitem{Bakshi0} G Bakshi, C Cao, Z Chen (1997). Empirical performance of alternative option pricing models. The Journal of finance 52, 2003--2049.

\bibitem{Bakshi} G. Bakshi, N. Kapadia, D.B. Madan (2003). ''Stock return characteristics, skew laws, and differential 
pricing of individual equity options'', Review of Financial Studies 16, 101--143.

\bibitem{Bakshi2} G. Bakshi, D. Madan, G. Panayotov (2010). Returns of Claims on the Upside and the Viability of U-shaped Pricing Kernels. Journal of Financial Economics 97, 130--154.



\bibitem{Bliss} R. Bliss, N. Panigirtzoglou (2004). ''Option-Implied Risk Aversion Estimates'' Journal of Finance 59, 407--446.

\bibitem{BreedenLitzenberger} D. T. Breeden, R. H. Litzenberger (1978). ''Prices of State-Contingent Claims Implicit in Option Prices'' The Journal of Business 51, 621--651.


\bibitem{Buchen} P.W. Buchen, M. Kelly (1996). ''The maximum entropy distribution of an asset inferred from option prices'' J. Financial and quantitative analysis 31, 143--159. 

\bibitem{Bondarenko} O. Bondarenko (2003). ''Estimation of risk-neutral densities using positive convolution approximation'' Journal of Econometrics 116, 85--112.


\bibitem{Brous}  Brous, P., Ince, U., Popova, I. : Volatility forecasting and liquidity: Evidence from individual stocks.
J Deriv Hedge Funds (2010) 16: 144--159.

\bibitem{Brunner_et_al} B. Brunner, R. Hafner (2003). ''Arbitrage-free estimation of the risk-neutral density from 
implied volatility smile'' The Journal of Computational Finance 7, 75--106.

  
\bibitem{Carr} P. Carr, K. Ellis, V. Gupta (1998). ''Static Hedging of Exotic Options''  Journal of Finance 53, 1165--1190.

\bibitem{CarrWu1} P. Carr, L. Wu (2009). ''Variance Risk Premiums''  Rev. Financ. Stud. 22, 1311--1341. 

\bibitem{CarrWu2}
P. Carr, L. Wu (2016). ''Analyzing volatility risk and risk premium in option contracts: A new theory'' 
Journal of Financial Economics 120, 1--20.

\bibitem{chalamandris}
G. Chalamandris, L. Rompolis (2012). ''Exploring the role of the realized return distribution in the formation of the
implied volatility smile'' J. Banking and Finance 36 , 1028--1044.


\bibitem{chernov}
M. Chernov, E. Ghysels (2000). ''A study towards a unified appraoch to the joint estimation of objective and risk neutral 
measures for the purpose of options valuation'' J. Financ. Econ. 56,  407--458.


\bibitem{Conrad}
J. Conrad, R.F. Dittmar, E. Ghysels (2013), ''Ex Ante Skewness and Expected Stock Returns'', The Journal of Finance 
LXVIII, 85--124.

\bibitem{corrado}
C.J. Corrado, T. Su (1997). ''Implied volatility skews and stock return skewness and kurtosis implied by stock option 
prices'' Eur. J. Finance 3 , 73--85.

\bibitem{corrado0}
C. Corrado (2007). ''The hidden martingale restriction in Grahm-Charlier option prices'' 
The Journal of Futures Markets 27, 517--534. 

\bibitem{Derman} E, Derman, D. Ergener, I. Kani (1995). ''Static Options Replication'' Journal of Derivatives 2, 78--95.

\bibitem{EngleFiglewski}
R. Engle, S. Figlewski (2015). ''Modeling the Dynamics of Correlations among Implied Volatilities'' Review of Finance 19, 991--1018. 

\bibitem{Fama}
E.F. Fama (1963). ''Mandelbrot and the Stable Paretian Hypothesis'' The Journal of Business 36, 420--429.

\bibitem{Fengler} M. Fengler (2005). ''Semiparametric modeling of Implied Volatility'', Springer.



\bibitem{Figlewski} 
S. Figlewski (2010). Estimating the Implied Risk-Neutral Density for the US Market Portfolio. In
T. Bollerslev, J. Russell, \& M. Watson (Eds.), Volatility and Time Series Econometrics: Essays
in Honor of Robert Engle chapter 15, (pp. 323–353). Oxford University Press.


\bibitem{halperin} I. Halperin, A. Itkin (2014). ''Pricing options on illiquid assets with liquid proxies using utility indifference and dynamic-static hedging'' Quant. Fin. 14,  427--442.

\bibitem{Heston} S. Heston (1993). "A Closed-Form Solution for Options with Stochastic Volatility with Applications to Bond and Currency Options". The Review of Financial Studies. 6, 327--343.

\bibitem{hocquard} A. Hocquard, N. Papapgeorgiou, B. Remillard (2012). ''The payoff distribution model: an application to dynamic portfolio insurance'' Quant. Fin. 15, 299--312.


\bibitem{Jackwerth} J.C. Jackwerth, M. Rubinstein (1996). ''Recovering probability distributions from option prices'' 
The Journal of Finance 51, 1611--1631.



\bibitem{jarrow-rudd}
R. Jarrow, A. Rudd (1982). ''Approximate option valuation for arbitrary stochastic processes'' J. Financ. Econ. 10, 347--369.
 
\bibitem{JPR}
E. Jondeau, S.H. Poon, M. Rockinger (2007). ''Financial Modeling under Non-Gaussian Distributions'' Springer. 

\bibitem{Kristensen}
D. Kristensen, A. Mele (2011). Adding and subtracting Black-Scholes: A new approach to approximating derivative prices in continuous-time models, Journal of Financial Economics, 102, 390--415.


\bibitem{longstaff}
F. Longstaff (1995) ''Option pricing and the martingale restriction'' Rev. Financ. Stud. 8, 1091--1124.
  
  
\bibitem{MCC}
D. Madan, P. Carr, E. Chang (1998). ''The Variance Gamma Process and Option Pricing'' European Finance Review 2, 79--105.
  
  
\bibitem{Mandelbrot} B. Mandelbrot (1963). ''The Variation of Certain Speculative Prices'' The Journal of Business 36, 394--419. 
  
\bibitem{Melik} W.R. Melik, C.P. Thomas (1997). Recovering an Asset Implied PDF from Option Prices: An Application to Crude Oil during the Gulf Crisis. Journal of Financial and Quantitative Analysis, 32, 91--115.  
  
\bibitem{Merton} R.C. Merton (1973). "Theory of Rational Option Pricing". Bell Journal of Economics and Management Science. The RAND Corporation. 4, 141--183.




\bibitem{Ross} S. Ross (2015).  ''The Recovery Theorem'' The Journal of Finance 70, 615--648.


\bibitem{Rubinstein} M. Rubinstein (1994). ''Implied Binomial Trees'' The Journal of Finance 49, 771--818.

\bibitem{Rubinstein2} M. Rubinstein (1998). ''Edgeworth binomial trees'' The Journal of Derivatives 5, 20--27.


\bibitem{SX}
Song, Z., Xiu, D. : A tale of two option markets: Pricing kernels and volatility risk, Journal of Econometrics, 190, 176--196
(2016)

\bibitem{Stutzer} M. Stutzer (1996). ''A simple nonparametric approach to derivative security valuation''  The Journal of Finance 51, 1633--1652.

\bibitem{SEJD} Szakmary, A., Evren, O., Jin, K. K., Davidson, W.N. (2003). The predictive power of implied volatility: Evidence from 35 futures markets. Journal of Banking \& Finance, 27, 2151--2175. 

\bibitem{Talponen} Talponen J. (2018), Matching distributions: Derivatives pricing with physical density shape correction.
ArXiv Preprint. 

\end{thebibliography}





\end{document}